\documentclass{aa}
\usepackage{graphicx}
\usepackage{txfonts}
\usepackage{natbib}
\usepackage{color}
\usepackage[colorlinks, allcolors=blue]{hyperref}
\newcommand{\HRI}{HRI$_\mathrm{EUV}$}

\begin{document}

\title{EUV brightenings in the quiet-Sun: Signatures in spectral and\\ imaging data from the Interface Region Imaging Spectrograph}

\author{C. J. Nelson$^{1}$\thanks{ESA Research Fellow}, F. Auch\`ere$^{2}$, R. Aznar Cuadrado$^{3}$, K. Barczynski$^{4,5}$, E. Buchlin$^{2}$, L. Harra$^{5,4}$, D. M. Long$^{6,7}$, S. Parenti$^{2}$, H. Peter$^{3}$, U. Sch\"uhle$^{3}$, C. Schwanitz$^{5,4}$, P. Smith$^{6}$, L. Teriaca$^{3}$, C. Verbeeck$^{8}$, A. N. Zhukov$^{8,9}$, D. Berghmans$^{8}$}

\offprints{chris.nelson@esa.int}
\institute{$^1$European Space Agency (ESA), European Space Research and Technology Centre (ESTEC), Keplerlaan 1, 2201 AZ, Noordwijk, The Netherlands.\\
$^2$Universit\'e Paris-Saclay, CNRS, Institut d’astrophysique spatiale, 91405, Orsay, France.\\
$^{3}$Max Planck Institute for Solar System Research, Justus-von-Liebig-Weg 3, 37077 Göttingen, Germany.\\
$^{4}$ETH-Zurich, Wolfgang-Pauli-Str. 27, 8093 Zurich, Switzerland.\\
$^{5}$Physikalisch-Meteorologisches Observatorium Davos, World Radiation Center, 7260, Davos Dorf, Switzerland.\\
$^{6}$UCL-Mullard Space Science Laboratory, Holmbury St. Mary, Dorking, Surrey, RH5 6NT, UK.\\
$^{7}$Astrophysics Research Centre, School of Mathematics and Physics, Queen's University Belfast, Belfast BT7 1NN, UK\\
$^{8}$Solar-Terrestrial Centre of Excellence – SIDC, Royal Observatory of Belgium, Ringlaan -3- Av. Circulaire, 1180 Brussels, Belgium.\\
$^{9}$Skobeltsyn Institute of Nuclear Physics, Moscow State University, 119992 Moscow, Russia.}

\date{}

\abstract
{Localised transient EUV brightenings, sometimes named `campfires', occur throughout the quiet-Sun. However, there are still many open questions about such events, in particular regarding their temperature range and dynamics.}
{In this article, we aim to determine whether any transition region response can be detected for small-scale EUV brightenings and, if so, to identify whether the measured spectra correspond to any previously reported bursts in the transition region, such as Explosive Events (EEs).}
{EUV brightenings were detected in a $\sim29.4$ minute dataset sampled by Solar Orbiter's Extreme Ultraviolet Imager (EUI) on $8$ March $2022$ using an automated detection algorithm. Any potential transition region response was inferred through analysis of imaging and spectral data sampled through coordinated observations conducted by the Interface Region Imaging Spectrograph (IRIS).}
{EUV brightenings display a range of responses in IRIS slit-jaw imager (SJI) data. Some events have clear signatures in the \ion{Mg}{II} and \ion{Si}{IV} SJI filters, whilst others have no discernible counterpart. Both extended and more complex EUV brightenings are found to, sometimes, have responses in IRIS SJI data. Examples of EUI intensities peaking before, during, and after their IRIS counterparts were found in lightcurves constructed co-spatial to EUV brightenings. Importantly, therefore, it is likely that not all EUV brightenings are driven in the same way, with some events seemingly being magnetic reconnection driven and others not. A single EUV brightening occurred co-spatial to the IRIS slit, with the returned spectra matching the properties of EEs.}
{EUV brightenings is a term used to describe a range of small-scale event in the solar corona. The physics responsible for all EUV brightenings is likely not the same and, therefore, more research is required to assess their importance towards global questions in the field, such as coronal heating.}

\keywords{Sun: activity; Sun: atmosphere; Sun: transition region; Sun: UV radiation}
\authorrunning{Nelson et al.}
\titlerunning{Transition Region Response To EUV brightenings}

\maketitle

\section{Introduction}
\label{Introduction}

Brightening events with sub-Mm scales and lifetimes of the order of minutes are seemingly ubiquitous throughout the solar atmosphere, from the photosphere to the corona, from the quiet-Sun to active regions. Such events were first detected more than one century ago by \citet{Ellerman17}, who identified transient intensity increases, now known as Ellerman bombs (EBs), in the wings of H$\alpha$ line profiles sampled within an active region. Since then, a plethora of other classes of brightenings have been reported in the literature including, but not limited to, Explosive Events (EEs; \citealt{Brueckner83}), blinkers (\citealt{Harrison97}), and UV bursts (\citealt{Young18}), each with their own unique properties. The recent launches of satellites such as the Interface Region Imaging Spectrograph (IRIS; \citealt{DePontieu14}) and Solar Orbiter (\citealt{Muller20}) have opened a new window into localised brightenings in the solar atmosphere, allowing us to probe the transition region and corona on unprecedented scales. These observational developments have led to improved understanding of many events such as IRIS bursts (\citealt{Peter14}) and EUV brightenings (\citealt{Berghmans21}), thereby improving our understanding of energy release in the upper solar atmosphere. Importantly, coordinated observations between these two satellites has provided the opportunity to understand linkages between brightenings identified in different temperature windows, offering insights into the complex physics responsible for these events and their potential role in coronal heating (\citealt{Cargill15}).

The detection of thousands of small-scale EUV brightenings (\citealt{Berghmans21}), sometimes referred to as `campfires', in quiet-Sun imaging data was one of the key early results of Solar Orbiter's Extreme Ultraviolet Imager (EUI; \citealt{Rochus20}). These EUV brightenings have been shown to occur right down to the spatial resolution of EUI data sampled using the $17.4$ nm High-Resolution Imager (\HRI) telescope, at scales below what could have been routinely detected in data from the Solar Dynamics Observatory's Atmospheric Imaging Assembly (SDO/AIA; \citealt{Lemen12}). Despite their recent discovery, these small-scale EUV brightenings have been widely studied by the community through both observations and numerical simulations. \citet{Zhukov21}, for example, combined imaging data sampled by SDO/AIA with data from \HRI\ whilst Solar Orbiter was away from the Sun-Earth line. Those authors used the distinct viewing angles of the two satellites to conduct stereoscopy of these events, finding that they typically form at heights between $1$ Mm and $5$ Mm in the solar atmosphere. Additionally, analysis of the line-of-sight photospheric magnetic field sampled by SDO's Helioseismic and Magnetic Imager (SDO/HMI; \citealt{Scherrer12}) by \citet{Panesar21} and Solar Orbiter's Polarimetric and Helioseismic Imager (PHI; \citealt{Solanki20}) by \citet{Kahil22} has shown that many (potentially around $70$ \%) of these localised EUV brightenings occur co-spatial to bi-poles, where magnetic reconnection is thought to be possible. On top of this, both numerical simulations (\citealt{Chen21}) and magnetic field extrapolations (\citealt{Barczynski22}) have supported the assertion that magnetic reconnection in the upper solar atmosphere could be driving these events.

\begin{figure*}
\includegraphics[width=0.99\textwidth]{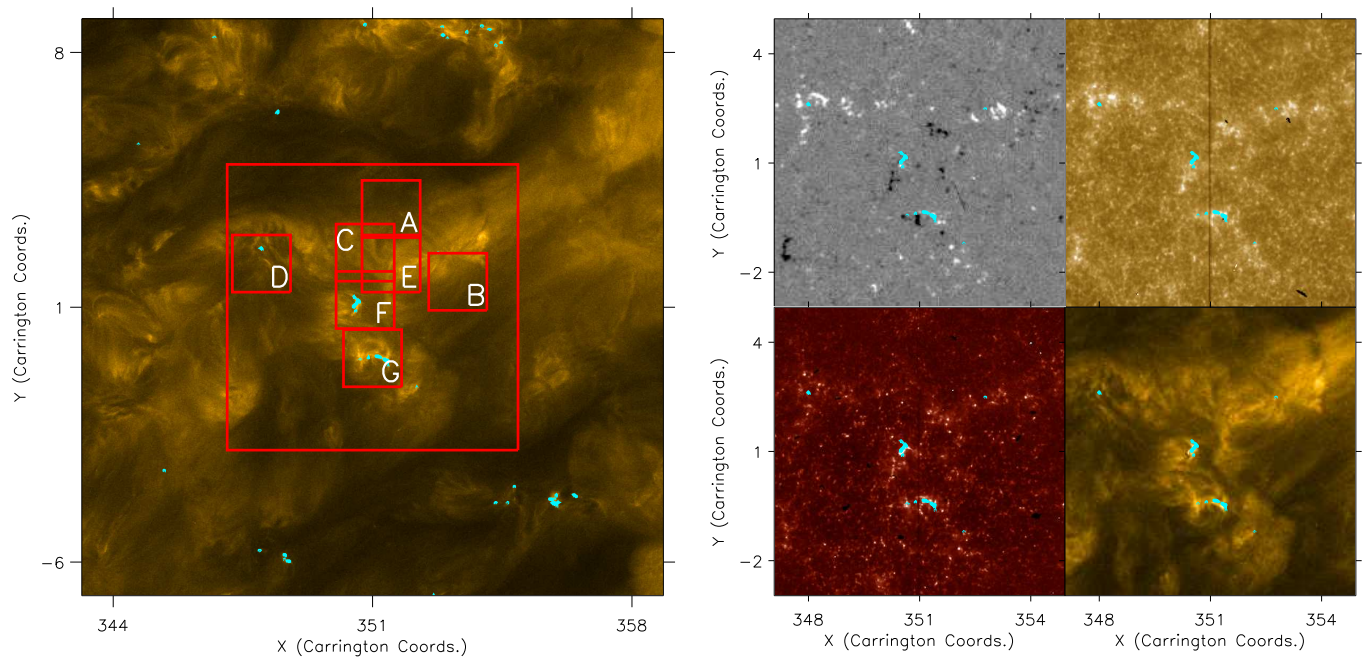}
\caption{Context image for the quiet-Sun region studied in this article. In the left-hand panel, we plot a zoom-in of the \HRI\ FOV observed at $00$:$04$:$21$ UT on $8$ March $2022$. The larger red over-laid box outlines the IRIS FOV plotted in the right-hand panels, whilst the smaller, labelled red boxes (corresponding to approximately $19.1$ Mm $\times~19.1$ Mm) outline the sub-FOVs which contained the EUV brightening activity studied here. In the right-hand panels, we plot the IRIS FOV as sampled by SDO/HMI line-of-sight magnetograms (top left), the IRIS \ion{Mg}{II} $279.6$ nm filter (top right), the IRIS \ion{Si}{IV} $140$ nm filter (bottom left), and the SDO/AIA $17.1$ nm filter (bottom right). The cyan contours indicate the locations of EUV brightenings identified in the plotted \HRI\ frame.}
\label{CFs_IRIS_Context}
\end{figure*}

One research area which is relatively unexplored relating to EUV brightenings identified in EUI data, is whether they display spectral signatures across other temperature regimes of the atmosphere (e.g. in the chromosphere and transition region). Using imaging data sampled by SDO/AIA, both \citet{Panesar21} and \citet{Dolliou23} were able to identify that cool plasma was present in the lower solar corona co-spatial to some EUV brightenings, whilst \citet{Huang23} studied the spectral response of three events in SPectral Imaging of the Coronal Environment (SPICE; \citealt{SPICE20}) data finding responses in cooler lines for all studied events. However, whether a response to this plasma would be detected in the transition region for all events is currently unknown. If these EUV brightenings are driven by magnetic reconnection, then it is certainly plausible that they could form co-spatial and co-temporal to other burst signatures identified in observations sampling different temperature windows in the solar atmosphere. In the quiet-Sun, this could include events such as photospheric Quiet-Sun Ellerman-like Brightenings (QSEBS; \citealt{Rouppe16,Nelson17}) or transition region EEs (see \citealt{Brueckner83, Huang19}). The coordinated observations conducted by \HRI\ and IRIS during the first science orbit of Solar Orbiter (the EUI data are summarised in \citealt{Berghmans23}) offer an ideal opportunity to assess the transition region response to these EUV brightenings and, therefore, potentially better understand the effects of magnetic reconnection in the solar atmosphere.

In this article, we aim to further our understanding of small-scale EUV brightenings through analysis of data collected during coordinated observations between \HRI\ and IRIS on $8$ March $2022$. Specifically, we aim to investigate whether signatures of these EUV brightenings are present in spectral and imaging data sampling transition region temperatures and, if so, whether these signatures match those of other well-known bursts (e.g. EEs). We structure our work as follows: We introduce the data analysed in this article in Sect.~\ref{Observations}; The results of our analysis are then presented in Sect.~\ref{Results}; A brief discussion is included in Sect.~\ref{Discussion}; Before we draw our conclusions in Sect.~\ref{Conclusions}.

\section{Observations}
\label{Observations}

In this article, we analyse a region of the quiet-Sun observed by \HRI\ using the $17.4$ nm filter between $00$:$04$:$12$ UT and $00$:$33$:$33$ UT (all times have been converted to the reference frame of Earth) on $8$ March $2022$\footnote{EUI data release 5.0: https://doi.org/10.24414/2qfw-tr95}. At $00$:$00$:$20$ UT, the sub-solar point of Earth was $1.2^\circ$ East and $-3.0^\circ$ South of the sub-solar point of Solar Orbiter, allowing coordination with Earth-bound observational infrastructure. The $2048$ pixel $\times\ 2048$ pixel images collected by \HRI\ have a pixel scale of $0.492$\arcsec\ which, as Solar Orbiter was at a distance of approximately $0.49$ AU from the Sun during this experiment, corresponds to approximately $174$ km on the solar surface at disk centre. The cadence of images within this dataset was $3$ s, while the exposure time was $1.65$ s. We downloaded level-2 data from the Solar Orbiter Archive, with the identification of EUV brightenings being accomplished by employing the automated algorithm developed by \citet{Berghmans21}. This algorithm converted the original \HRI\ data into Carrington coordinates ($1$ Carrington degree is equal to approximately $12.1$ Mm on the solar surface), which involved a remapping of the images onto a $2172$ pixel $\times~2172$ pixel grid whilst maintaining the spatial sampling of each image. Following this, the properties of small-scale EUV brightenings distinct from the shot-noise were extracted. Here, we analyse seven regions within the FOVs of both \HRI\ and IRIS which were found to contain EUV brightening activity with sufficient areas and durations such that meaningful comparisons could be made between the instruments.

\begin{figure*}
\includegraphics[width=0.99\textwidth]{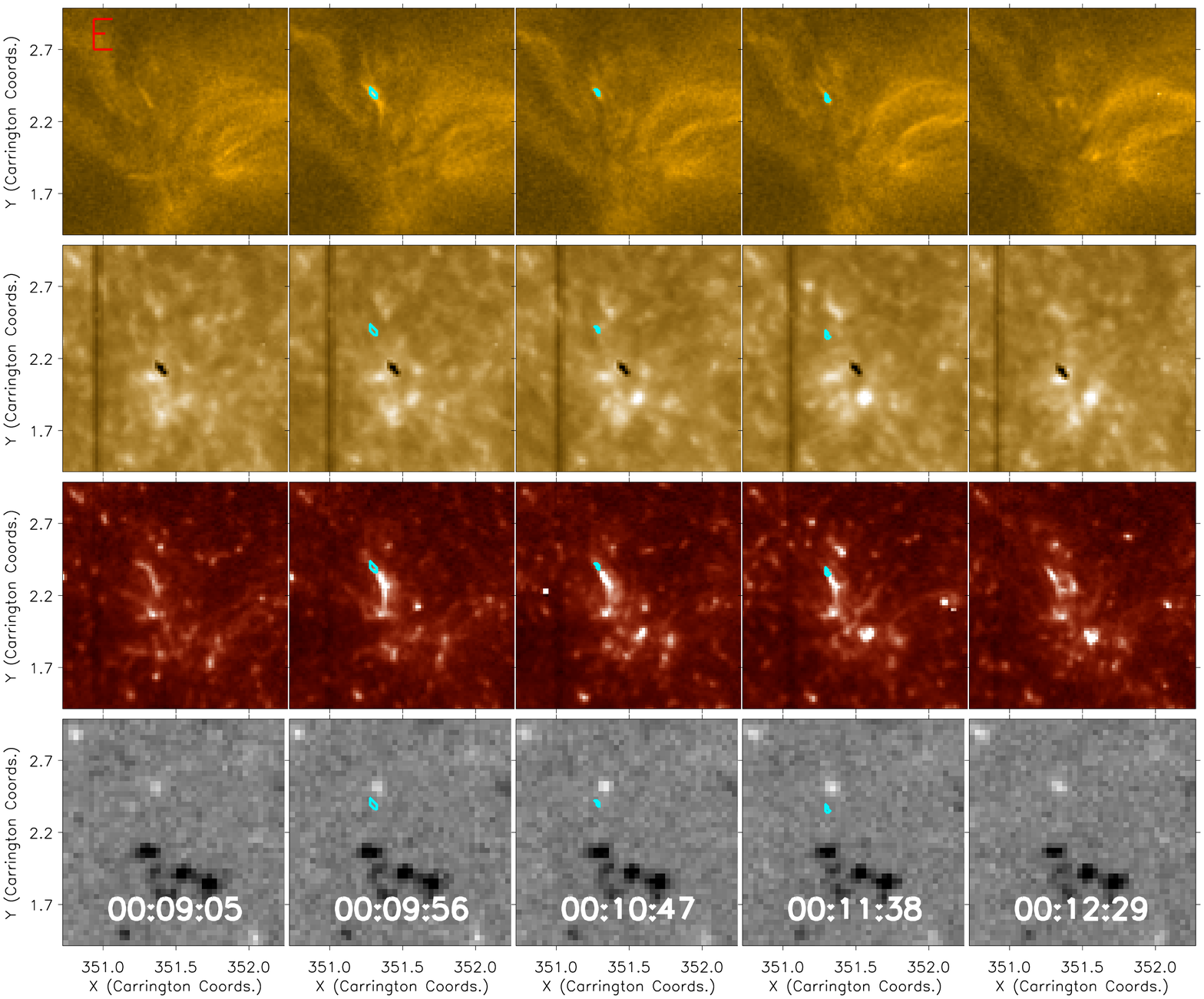}
\caption{Example of an extended EUV brightening from region `E'. The top row plots the \HRI\ intensity at five time-steps over the course of around $204$ s with the EUV brightening activity being outlined by the cyan contours. The second row plots the co-spatial and co-temporal IRIS \ion{Mg}{II} $279.6$ nm filter SJI response (the dark spot in the centre of the images is a known blemish in the filter), while the third row plots the equivalent but for the \ion{Si}{IV} $140$ nm filter. A clear brightening is evident in the \ion{Si}{IV} $140$ nm channel with a similar shape to, but extending slightly further to the south from, the EUV brightening in the \HRI\ data. The bottom row plots the line-of-sight photospheric magnetic field at this location as sampled by the SDO/HMI instrument. An animation of this region is available with the online version of this article.}
\label{CFs_IRIS_E1}
\end{figure*}

\begin{figure*}
\includegraphics[width=0.99\textwidth]{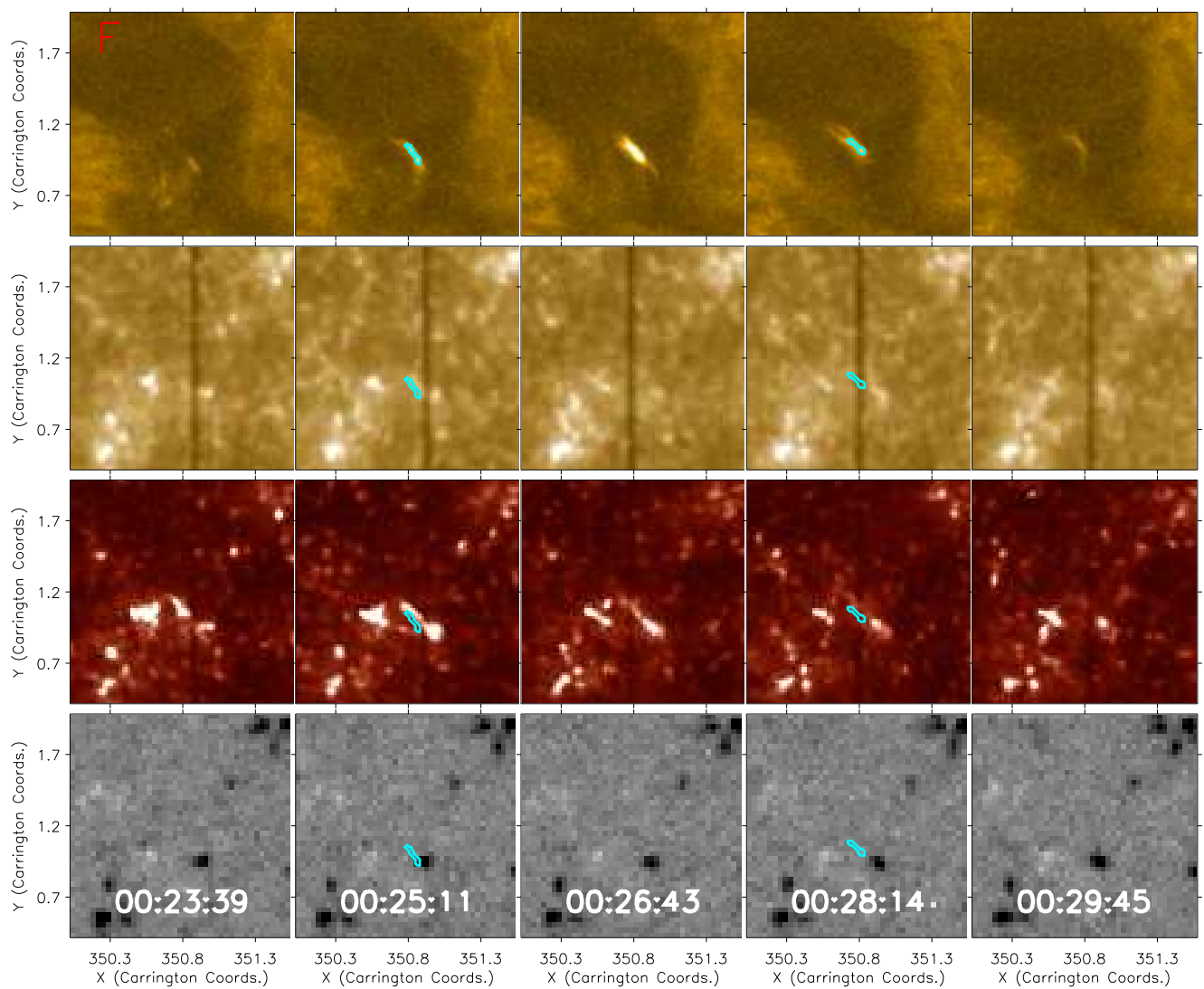}
\caption{Same as Fig.~\ref{CFs_IRIS_E1} but for an extended EUV brightening identified in region `F'. No large off-set (a small north-south off-set may be present) is apparent between this EUV brightening and the co-temporal IRIS \ion{Si}{IV} $140$ nm SJI response. The IRIS slit passed through this event during its occurrence, as can be seen clearly in the fourth column. An animation of this region is available with the online version of this article.}
\label{CFs_IRIS_F4}
\end{figure*}

\begin{figure*}
\includegraphics[width=0.99\textwidth]{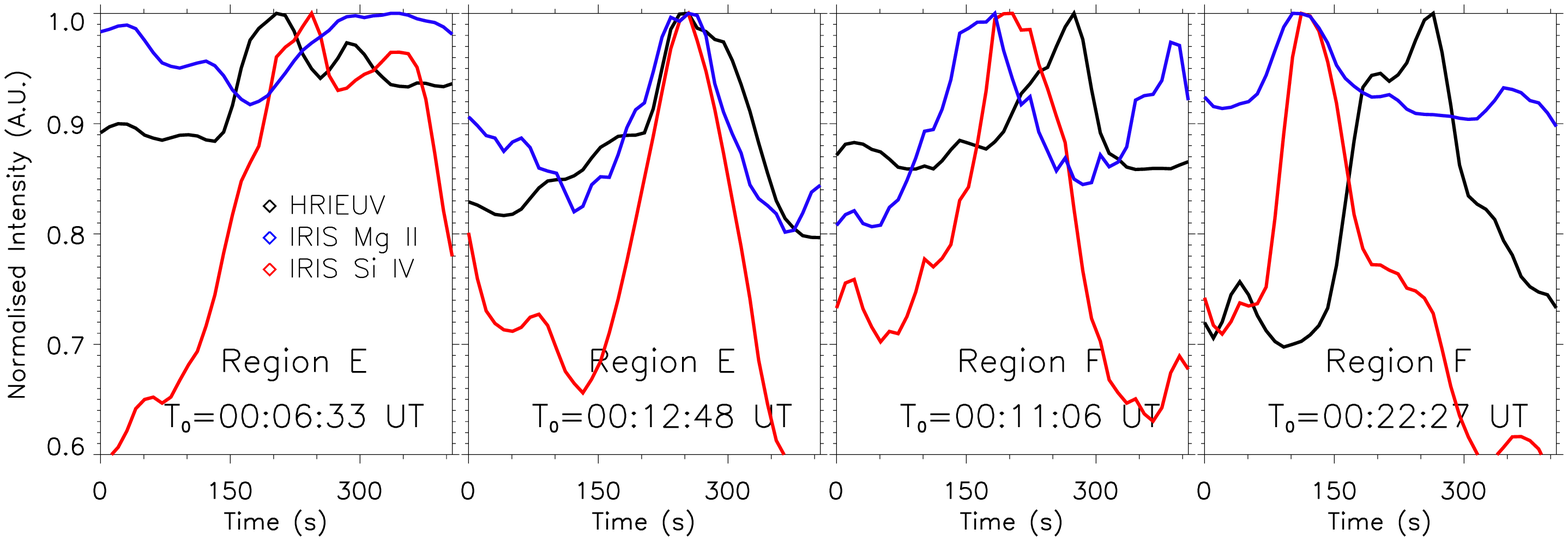}
\caption{Lightcurves of the mean intensity around four extended EUV brightening within regions E and F. The black lines plot lightcurves constructed from the \HRI\ data, while the blue and red lines plot lightcurves constructed using the IRIS \ion{Mg}{II} $279.6$ nm and \ion{Si}{IV} $140$ nm SJI filters, respectively. The left two panels plot lightcurves constructed from small boxes (the sizes of which were close to $1.8$ Mm$^2$, with the exact dimensions being varied to encompass the different orientations of individual EUV brightenings and any co-spatial associated brightenings in IRIS data) around two EUV brightenings within region E, with the first panel corresponding to the event plotted in Fig.~\ref{CFs_IRIS_E1}. The right two panels plot lightcurves constructed from small boxes (like those constructed for region E) around two EUV brightenings within region F, with the fourth panel corresponding to the event plotted in Fig.~\ref{CFs_IRIS_F4}. The \HRI\ intensity is found to peak before (first panel), co-temporally with (second panel), and after (third and fourth panels) the IRIS \ion{Si}{IV} SJI $140$ nm SJI filter intensity for different events. Clear brightening events are also found in the \ion{Mg}{II} $279.6$ nm filters for the right-most three of these four events. The time defined as $t=0$ s is over-laid on each panel.}
\label{CFs_IRIS_lightcurves}
\end{figure*}

Whether any transition region response to the EUV brightenings could be detected was investigated using data sampled by the IRIS satellite. Between $00$:$04$:$20$ UT and $00$:$34$:$09$ UT, IRIS ran a dense $8$-step raster with slit-jaw imager (SJI) intensity maps sampled by the $279.6$ nm \ion{Mg}{II} (temperature of $\sim2\times10^{4}$\ K) and $140$ nm\ \ion{Si}{IV} (temperature of $\sim8\times10^{4}$\ K) filters in sequence. These SJI data covered a FOV of $169$\arcsec$\times175$\arcsec\ centred at heliographic coordinates of $x_\mathrm{c}=-41$\arcsec, $y_\mathrm{c}=126$\arcsec, with pixel scales of $0.33$\arcsec\ (horizontal scale of $239$ km) and cadences of $10$ s. In addition to the SJI data, IRIS sampled spectra around several lines in the near- and far-ultraviolet including \ion{C}{II} $133.7$ nm ($\sim2\times10^{4}$\ K), \ion{Si}{IV} $139.4$ nm ($\sim8\times10^{4}$\ K), \ion{Si}{IV} $140.3$ nm ($\sim8\times10^{4}$\ K), and \ion{Mg}{II} k $279.6$ nm ($\sim10^{4}$\ K). The raster data recorded for each spectral window had a spectral sampling of $0.005$ nm, a raster step-size of $0.35$\arcsec\ (horizontal scale of $254$ km), and a pixel scale along the slit of $0.33$\arcsec. The time taken to sample the spectra at one raster position was $5$ s, giving a total raster cadence of approximately $40$ s and a total number of raster repeats of $44$ during the experiment. The OBSID for this observational campaign was $3600607428$.

Finally, we also studied co-spatial imaging and line-of-sight magnetic field data sampled by the SDO/AIA and SDO/HMI instruments, respectively. Our analysis of SDO/AIA data was limited to the $17.1$ nm filter, which was primarily used for alignment purposes. These data have pixel scales of $0.6$\arcsec\ ($435$ km) and cadences of $12$ s. The line-of-sight magnetic field maps collected by the SDO/HMI instrument have post-reduction pixel scales of $0.6$\arcsec\ ($435$ km) and cadences of $45$ s. Basic alignment between instruments was conducted by converting the FOVs observed by \HRI, IRIS, SDO/AIA, and SDO/HMI into Carrington coordinates at each time-step using the meta-data of each file. Closer alignments were completed by aligning local \HRI\ and SDO/AIA $17.1$ nm intensity maps around each studied EUV brightening as required. In Fig.~\ref{CFs_IRIS_Context}, we plot the FOV studied here as sampled by each of these four instruments. The left-hand panel plots an \HRI\ image for a region around the IRIS FOV, which is indicated by the larger over-laid red box. The seven smaller, red boxes labelled A-G indicate the sub-FOVs which contained the EUV brightening activity analysed here. The right-hand panels plot the region within the IRIS FOV (larger red box) for: The photospheric line-of-sight magnetic field inferred by the SDO/HMI instrument (top left); the intensity in the chromospheric IRIS \ion{Mg}{II} $279.6$ nm filter (top right); the intensity in the transition region IRIS $140$ nm \ion{Si}{IV} filter (bottom left); and the intensity in the SDO/AIA $17.1$ nm filter (bottom right). The locations of EUV brightenings identified in the plotted \HRI\ frame are outlined by the cyan contours. The similarities between features in the \HRI\ and SDO/AIA data indicates that our basic alignment is reasonable for general analysis. If one were, instead, to investigate the one-to-one relationship between intensities at pixels within different instruments, then closer alignment and interpolations would be required.

\section{Results}
\label{Results} 

\subsection{Elongated EUV brightenings with clear IRIS responses}

\subsubsection{Imaging response}

We begin our analysis by investigating whether any chromospheric or transition region signatures could be detected co-spatial to EUV brightenings which were elongated and appeared to extend along a specific path (behaviour that could potentially be described as `jet'-like) returned by the algorithm from \HRI\ data. We identified seven events within regions `D'-`F' that matched this description, with durations of the order several minutes. We note that this duration corresponds to the lifetime of any macro-structure containing EUV brightenings returned by the algorithm in the \HRI\ data, rather than individually detected EUV brightenings from the algorithm for reasons which will be discussed in the next paragraph. In Fig.~\ref{CFs_IRIS_E1}, we plot an example of an extended event which was found within region `E'. In the top row, we plot the intensity as sampled by \HRI\ over the course of around $204$ s. Initially, no EUV brightening is detected (first column) within this region before an event appears (cyan contour over-laid on the second column) which remains visible over the following $100$ s (third and fourth columns) before fading from view (fifth column). The second and third rows plot the co-spatial and co-temporal responses of the IRIS \ion{Mg}{II} $279.6$ nm and \ion{Si}{IV} $140$ nm filters. No clear signature is present co-spatial to the EUV brightening (if we assume a good alignment) in the \ion{Mg}{II} $279.6$ nm images, however, an extended structure is present in the \ion{Si}{IV} $140$ nm data. This localised event appears to extend slightly further south than the EUV brightening, and remains visible after the \HRI\ event is no longer apparent. No evidence of interactions, such as cancellation, between opposite polarity magnetic fields is present in the SDO/HMI line-of-sight photospheric magnetic field maps (bottom row).

Although an off-set is apparent between the transition region and coronal brightening activity presented in Fig.~\ref{CFs_IRIS_E1}, this is not a universal property. In fact, six of the seven extended EUV brightening events have IRIS \ion{Si}{IV} $140$ nm signatures seemingly perfectly co-spatial to the \HRI\ structures, whilst only one (the event plotted in Fig.~\ref{CFs_IRIS_E1}) is clearly off-set (beyond any slight alignment errors). In Fig.~\ref{CFs_IRIS_F4}, we plot an example of EUV brightening activity from region `F' which displays a transition region signature at the same spatial location over the course of around six minutes. Again, the top row plots the \HRI\ intensity at this location, with the cyan contours outlining the EUV brightening detected by the algorithm. The lack of an automatically detected EUV brightening in the third column, where a bright macro-structure is present, explains our decision to discuss regions of EUV brightening activity rather than individual events. Once again, no obvious brightening is apparent in the IRIS \ion{Mg}{II} $279.6$ nm data but a clear extended event is present in the \ion{Si}{IV} $140$ nm images. The transition region brightening appears to brighten slightly before the associated \HRI\ brightening activity. No clear bi-poles are present in the photospheric line-of-sight magnetic field at these locations (fourth row). As can be seen in the fourth column, the IRIS slit passed directly through this EUV brightening, meaning we are able to analyse its associated spectra later in this manuscript.

To further our analysis of these events, we also studied the temporal evolution of the intensities of extended EUV brightening activity in IRIS and \HRI\ data. In Fig.~\ref{CFs_IRIS_lightcurves}, we plot normalised lightcurves constructed by summing the intensity in small boxes (around $1.8$ Mm$^{2}$, with the exact dimensions varying depending on the topology of the feature) around four regions of extended EUV brightening activity, selected from `E' and `F', through time. The black lines plot the intensity sampled by \HRI\ with the EUV brightening activity being evident as intensity bumps of around $10$-$40$ \%\ of the pre-brightening intensity. The blue and red lines plot the co-spatial and co-temporal IRIS \ion{Mg}{II} $279.6$ nm and \ion{Si}{IV} $140$ nm intensities, respectively, for each event. Clear increases in intensity (up to $70$ \%) are apparent for each of these events in the \ion{Si}{IV} $140$ nm data, with the right-most three of these also showing \ion{Mg}{II} $279.6$ nm intensity increases (of up to $25$ \%). The temporal evolution of these lightcurves is not the same for each extended EUV brightening, with the \HRI\ brightening occurring either $50$ s before (first panel; corresponding to the event plotted in Fig.~\ref{CFs_IRIS_E1}), co-temporally with (second panel), or between $100$-$150$ s after (third and fourth panels; corresponding to the event plotted in Fig.~\ref{CFs_IRIS_F4}) the associated IRIS intensity increases. Lightcurves constructed from the remaining three regions of extended EUV brightening activity were less clear, due to lower intensity enhancements in both the \HRI\ and IRIS SJI filters, and as such are not plotted.

This fascinating range of results is difficult to explain through one model alone, with many sub-types of extended EUV brightenings appearing to be present. Here we shall discuss three specific examples. The first sub-type includes extended brightenings that are detected co-spatially and sequentially from cooler to hotter filters over the course of several minutes. Such signatures could correspond, for example, to gradual heating of plasma from chromospheric to coronal temperatures caused by a single instance of energy release in the solar atmosphere or the sequential release of energy at different layers of the solar atmosphere over this time-period. Other explanations are, of course, also possible. The second sub-type includes co-spatial extended brightenings where the intensity peaks co-temporally in all filters. Potential explanations for this include both a single instance of highly impulsive energy release heating relatively cool plasma to coronal temperatures almost instantaneously or the occurrence of rapid cooling which is detected in many different temperature bands almost instantaneously. The third sub-type includes extended brightenings where an off-set is apparent for the structure in different filters and where the intensity in the hotter channels peaks before the intensity in the lower temperature channels. This sub-type implies that not all EUV brightenings returned by the algorithm pin-point locations where heating is currently occurring in the solar atmosphere, with at least some events returned by the algorithm pin-pointing material that is in its cooling phase (presumably following an earlier heating event that may or may not have been detected depending on its properties). The spatial off-sets between the structure in different filters could indicate that this specific event was not magnetic reconnection driven at all, instead being similar in nature to the thermal instability driven supersonic downflows (caused by coronal rain) regularly observed in the transition regions of sunspots (e.g. \citealt{Nelson20, Nelson20b}). Movies depicting the temporal behaviour of the four regions studied in this sub-section are supplied with the online version of this article.

\subsubsection{Spectral response}

\begin{figure*}
\includegraphics[width=0.99\textwidth]{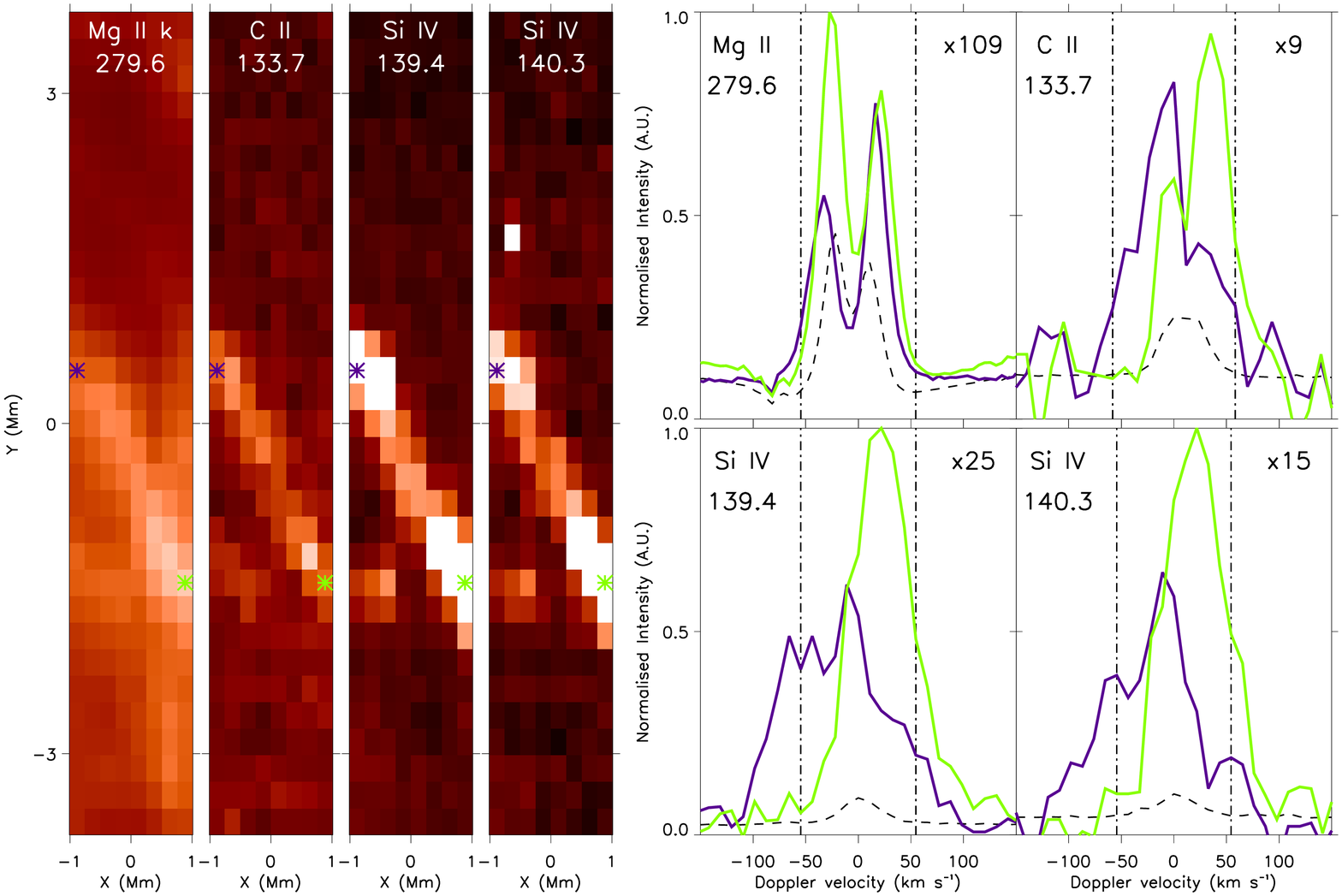}
\caption{Spectral response to the extended EUV brightening plotted in Fig.~\ref{CFs_IRIS_F4} as sampled at $00$:$25$:$21$ UT. The left four panels plot the integrated intensity over $110$ km s$^{-1}$ Doppler velocity windows around the rest wavelength for the \ion{Mg}{II} $279.6$ nm (first panel), \ion{C}{II} $133.7$ nm (second panel), \ion{Si}{IV} $139.4$ nm (third panel), and \ion{Si}{IV} $140.3$ nm (fourth panel) spectral lines. The four panels on the right plot the spectra sampled at the positions of the coloured crosses on the left hand panels (within the first and last raster positions). The black dashed lines plot the average spectra calculated from this raster, whilst the two vertical dot-dashed lines book-end the Doppler velocity windows integrated across to construct the left hand panels. Each spectral profile has been normalised against the peak value (DN/s) from these three spectra, with the normalisation factor printed in the top right corner. The spectra are slightly blue-shifted ($-22$ km s$^{-1}$; calculated from a single Gaussian fit to the \ion{Si}{IV} $139.4$ nm line) at the left side of the FOV (purple spectra), while they are slightly red-shifted ($+22$ km s$^{-1}$; calculated from a single Gaussian fit to the \ion{Si}{IV} $139.4$ nm line) at the right side of the FOV (green spectra).}
\label{CFs_IRIS_Spec}
\end{figure*}

\begin{figure}
\includegraphics[width=0.49\textwidth]{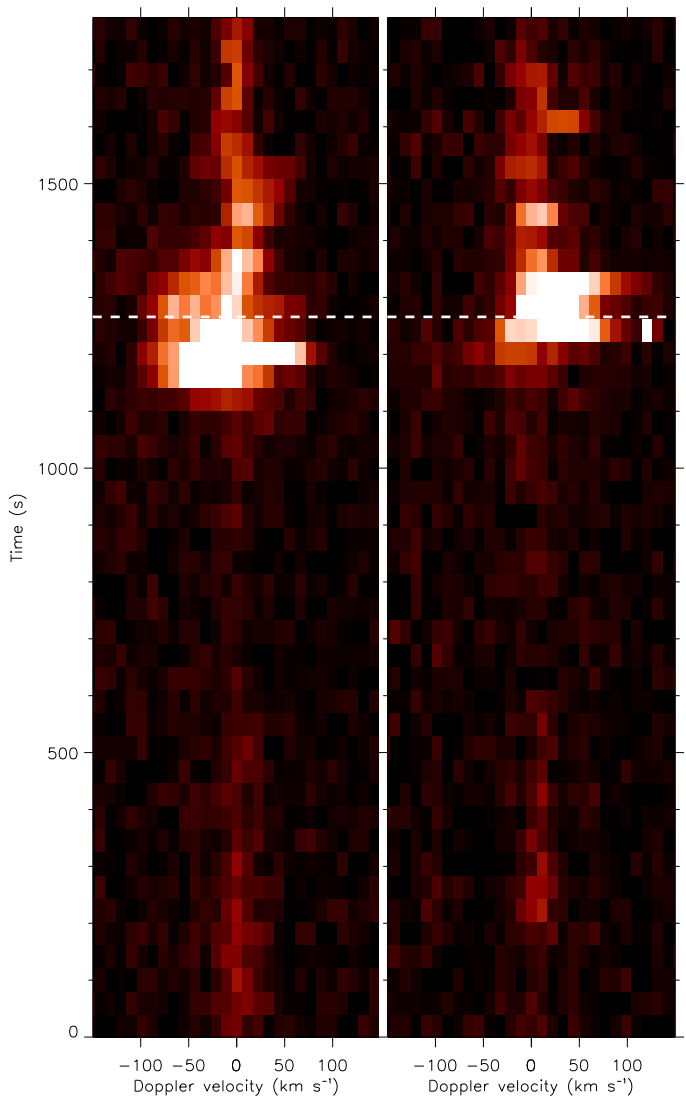}
\caption{Spectral-time maps constructed from the \ion{Si}{IV} $139.4$ nm line for the two pixels (left panel corresponds to the purple crosses while the right panel corresponds to the green crosses) plotted in Fig.~\ref{CFs_IRIS_Spec}. The dashed, white horizontal lines indicate the time-step plotted in that figure. The clear blue and red shifts measured at either end of the EUV brightening from the spectra are evident at multiple time-steps throughout the evolution of this apparent EEs.}
\label{CFs_IRIS_SpT}
\end{figure}

\begin{figure*}
\includegraphics[width=0.99\textwidth]{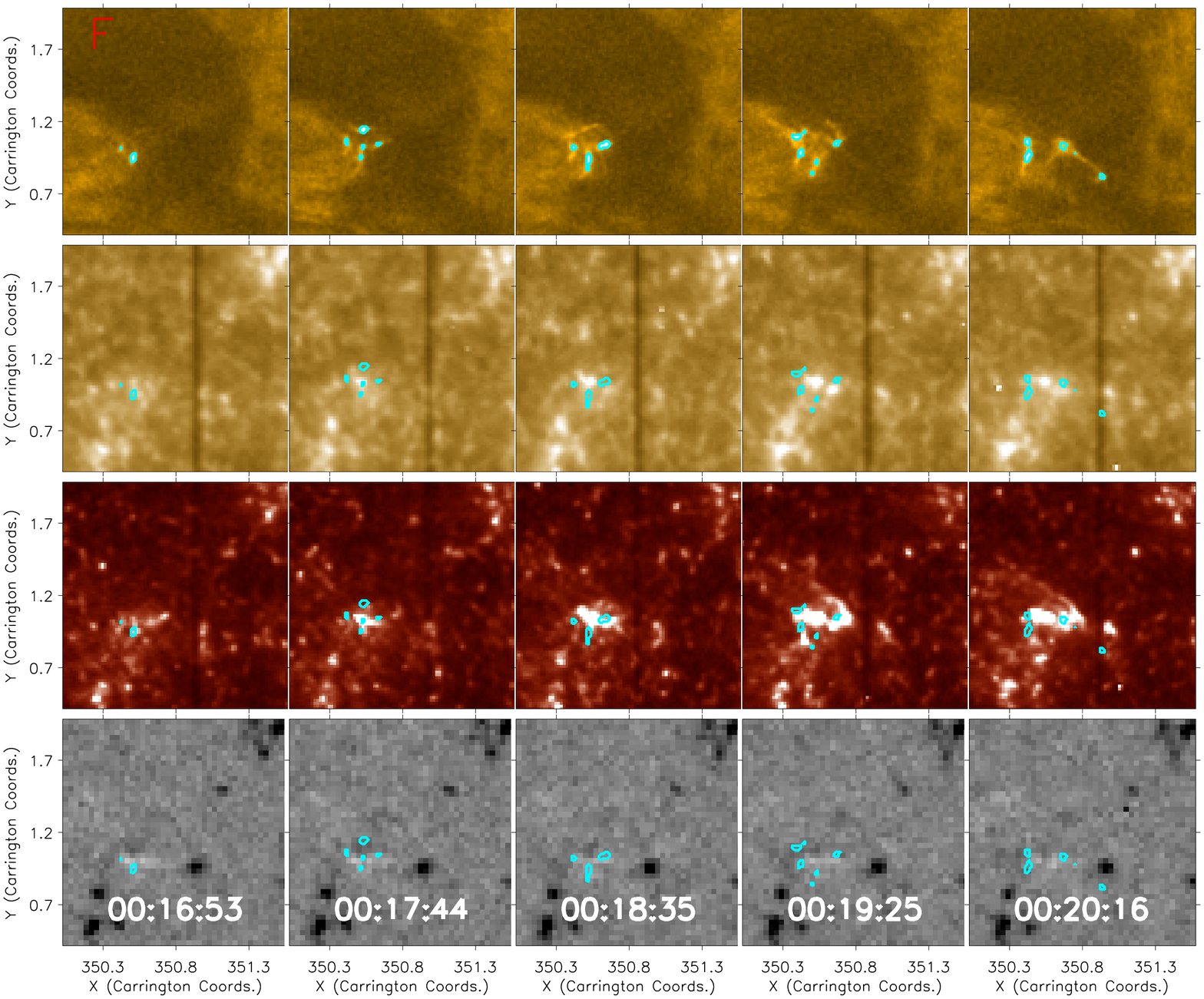}
\caption{Same as for Fig.~\ref{CFs_IRIS_F4} but for a more complex EUV brightening. Multiple small EUV brightenings are returned by the algorithm throughout this time in the \HRI\ images, outlined by the cyan contours. A small point-like brightening develops in the \ion{Mg}{II} channel co-spatial to the \HRI\ activity, whilst a well-defined structure is evident in the \ion{Si}{IV} filter. This structure includes two spiral arms, which are most clearly observed in the fourth column. An animation of this region is available with the online version of this article.}
\label{CFs_IRIS_F3}
\end{figure*}

To continue our investigation of extended EUV brightenings, we also analysed the spectra sampled co-spatial to the event plotted in Fig.~\ref{CFs_IRIS_F4}. In the left four panels of Fig.~\ref{CFs_IRIS_Spec}, we plot the summed intensity across a $110$ km s$^{-1}$ Doppler window centred on the rest wavelengths of four spectral lines sampled by IRIS at $00$:$25$:$21$ UT. A diffuse brightening is apparent in the \ion{Mg}{II} k $279.6$ nm image (first panel), however, this is not as clear as the long, thin brightening apparent in the \ion{C}{II} $133.7$ nm (second panel), \ion{Si}{IV} $139.4$ nm (third panel), and \ion{Si}{IV} $140.3$ nm (fourth panel) images. In the right-hand panels, we plot spectra sampled at the locations of the coloured crosses (where the colours of the crosses match the colours of the relevant lines) over-laid on the left-hand images, normalised against the peak DN/s values from the three lines plotted in each panel. The black dashed line on each panel plots the average `background' spectra across the raster scan. Clearly, the intensity is increased at the locations of the EUV brightening for each line, with the spectra also being broader. Clear oppositely directed Doppler shifts are also apparent at the two ends of the extended structure, with a velocity of $-22$ km s$^{-1}$ being measured at the left-hand edge of the raster (purple lines; from the \ion{Si}{IV} $140.3$ nm spectra using a single Gaussian fit) and a Doppler velocity of $+22$ km s$^{-1}$ being measured at the right-hand side of the raster (green lines). It is possible that these velocities correspond to reconnection out-flows from a central energy release location.

Studying the \ion{Si}{IV} spectral windows in more detail allowed us to make further inferences about this extended EUV brightening event. For example, the intensity ratio between the \ion{Si}{IV} lines was around $1.7$ (see the normalisation factors printed in the top right corner of each spectral panel in Fig~\ref{CFs_IRIS_Spec}) at both ends of the feature at multiple time-steps, indicating that this event may have been formed under optically thick conditions (see, for example, \citealt{Mathioudakis99}). Additionally, no over-laid absorption lines (e.g. \ion{Ni}{II}) were present on the \ion{Si}{IV} $139.4$ nm spectra indicating that this event was not an IRIS burst (the properties of which are described in, for example, \citealt{Peter14, Nelson22, Nelson22b}), which are only found in active regions (\citealt{Kleint22}). The lifetime, area, spectral shapes, and Doppler velocities of this event do, however, match well with the properties of EEs in the transition region (see, for example, \citealt{Teriaca04, Huang14, Huang19}).  Unfortunately, no density measurements were possible within this EUV brightening, as there was not enough signal to identify the \ion{O}{IV} $139.9$ nm line above the noise level. Future work studying data collected with longer exposure times would be required to investigate this further. 

In Fig.~\ref{CFs_IRIS_SpT}, we plot spectral-time diagrams for the two pixel locations studied in detail in Fig.~\ref{CFs_IRIS_Spec}. The white, dashed horizontal lines indicate the time-step plotted in that figure. The clear blue and red shifts at either end of the apparent EE are evident at multiple time-steps, with broadening of the spectra being apparent up to around $\pm100$ km s$^{-1}$. Several other EUV brightenings were observed to occur close to the IRIS slit during this dataset, however, this was the only example which displayed an unambiguous IRIS SJI response which could be traced through the spectra. Analysis of other features would require sub-pixel alignments between the two instruments across the FOV. We stress, therefore, that given the range of EUV brightening activity observed within this FOV, this result should not be interpreted as general to all such events. A larger, statistical study must be completed to assess general EUV brightening transition region spectra in the near future.

\subsection{Complex EUV brightenings with clear IRIS responses}

In addition to extended structures, several examples of more complex EUV brightening activity are present within regions `F'-`G'. For a distinction to the previous subsection, and because the connectivity of these events is difficult to infer from imaging data alone, we simply refer to these events as `complex'. In Fig.~\ref{CFs_IRIS_F3}, we plot an example of a region of complex EUV brightening activity within region `F' at five time-steps. Initially, two small point-like EUV brightenings are detected in the \HRI\ data (first panel of the first row), before this expands to be five small, distinct EUV brightenings in the second column (less than one minute later). After this, a period of rapid intensity increase occurs (third column) before several distinct spiral-like arms appeared (fourth column). Finally, this was followed by the occurrence of two extremely fast ($>300$ km s$^{-1}$) apparent jets propagating towards the top-left and bottom-right (the tip of this jet can be identified by the right-most EUV brightening detected by the algorithm in the fifth column) of the FOV. In the IRIS \ion{Mg}{II} $279.6$ nm channel (second row), a point-like brightening occurs (which is not morphologically similar to the features identified in the \HRI\ data) during the third and fourth columns at the same spatial location as the grouping of EUV brightenings. The \ion{Si}{IV} $140$ nm data (third row), on the other hand, displays similar spatial structuring as the \HRI\ data, with the spiral-like arms in the fourth column actually being clearer in the transition region image. The line-of-sight photospheric magnetic field does not display any evidence of cancellation, however, the coronal dynamics support the hypothesis that magnetic reconnection is driving this complex EUV brightening activity. The presence of many automatically detected EUV brightenings across a small FOV also suggests that we may be resolving localised instances of energy release within the larger-scale (potential) magnetic reconnection event. The dynamics discussed here can be more clearly seen in the movie included with the online version of this article.

\begin{figure}
\includegraphics[width=0.49\textwidth]{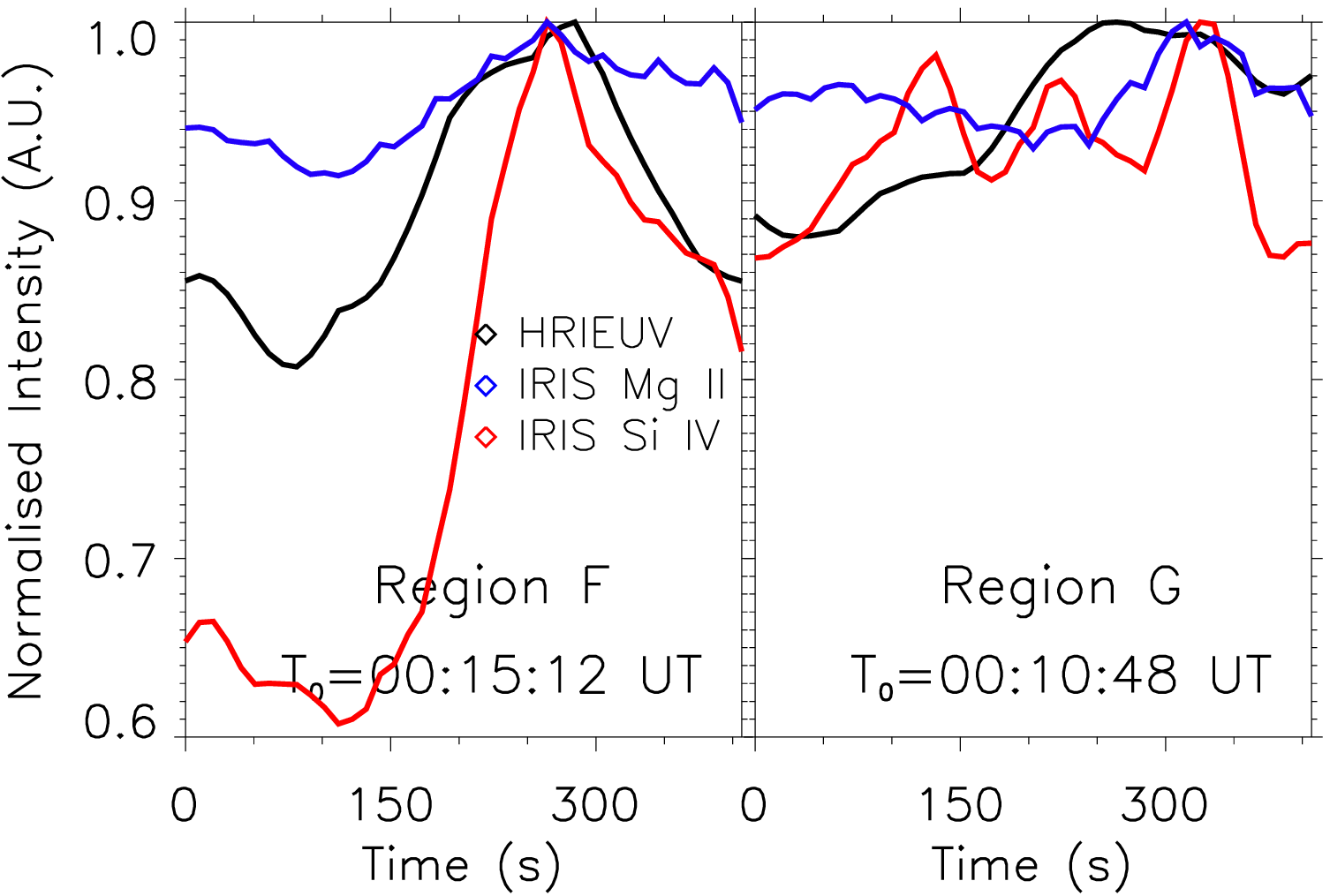}
\caption{Lightcurves of the mean intensity around two regions of complex EUV brightening activity within regions F (left panel) and G (right panel). Again, the black lines plot lightcurves constructed from the \HRI\ data, while the blue and red lines plot lightcurves constructed using the IRIS \ion{Mg}{II} $279.6$ nm and \ion{Si}{IV} $140$ nm SJI filters, respectively. The left panel plots the lightcurves constructed from a small box (similar to those discussed in Fig.~\ref{CFs_IRIS_lightcurves}) around the event plotted in Fig.~\ref{CFs_IRIS_F3}. The \HRI\ intensity is found to peak co-temporally with the increased emission in both IRIS SJI filters for the left panel. For the right panel, however, several short (lifetimes of the order seconds) bursts are evident in the IRIS \ion{Si}{IV} $140$ nm intensity during the occurrence of the EUV brightening. The time defined as $t=0$ s is over-laid on each panel.}
\label{CFs_IRIS_lightcurves_exp}
\end{figure}

Three further examples of complex EUV brightening activity which also displayed responses in the IRIS SJI data were present in these data. Two of these (one in each of regions F and G) were evident in the first \HRI\ frame meaning we were unable to study their full temporal evolutions, whilst the other occurred between $00$:$11$ UT and $00$:$17$ UT in region G. In Fig.~\ref{CFs_IRIS_lightcurves_exp}, we plot lightcurves for the two regions of complex EUV brightening activity for which their entire lifetimes were sampled. The left panel plots lightcurves constructed from a region (again around $1.8$ Mm$^2$ in area) surrounding the event plotted in Fig.~\ref{CFs_IRIS_F3} and displays a single intensity peak that is co-temporal in both the IRIS SJI filters and the \HRI\ data. The second region of complex EUV brightening activity displayed more variable behaviour, with multiple different intensity peaks being detected in the IRIS SJI \ion{Si}{IV} $140$ nm data (red line) over the course of around three minutes. Only a single intensity peak was evident in the IRIS SJI \ion{Mg}{II} $279.6$ nm filter (blue line) and the \HRI\ data (black line), with the \ion{Mg}{II} response occurring during the decay phase of the \HRI\ peak. Each of the intensity peaks in the IRIS SJI \ion{Si}{IV} $140$ nm filter manifests as small (diameters >$2$ Mm) bursts occurring at the same spatial location through time, with only one instance of brightening being apparent in the \HRI\ imaging data. Due to the complexity of this behaviour, it is not possible to offer direct insights into the exact physical mechanisms occurring at this location.

\subsection{EUV brightenings without clear IRIS responses}

\begin{figure*}
\includegraphics[width=0.99\textwidth]{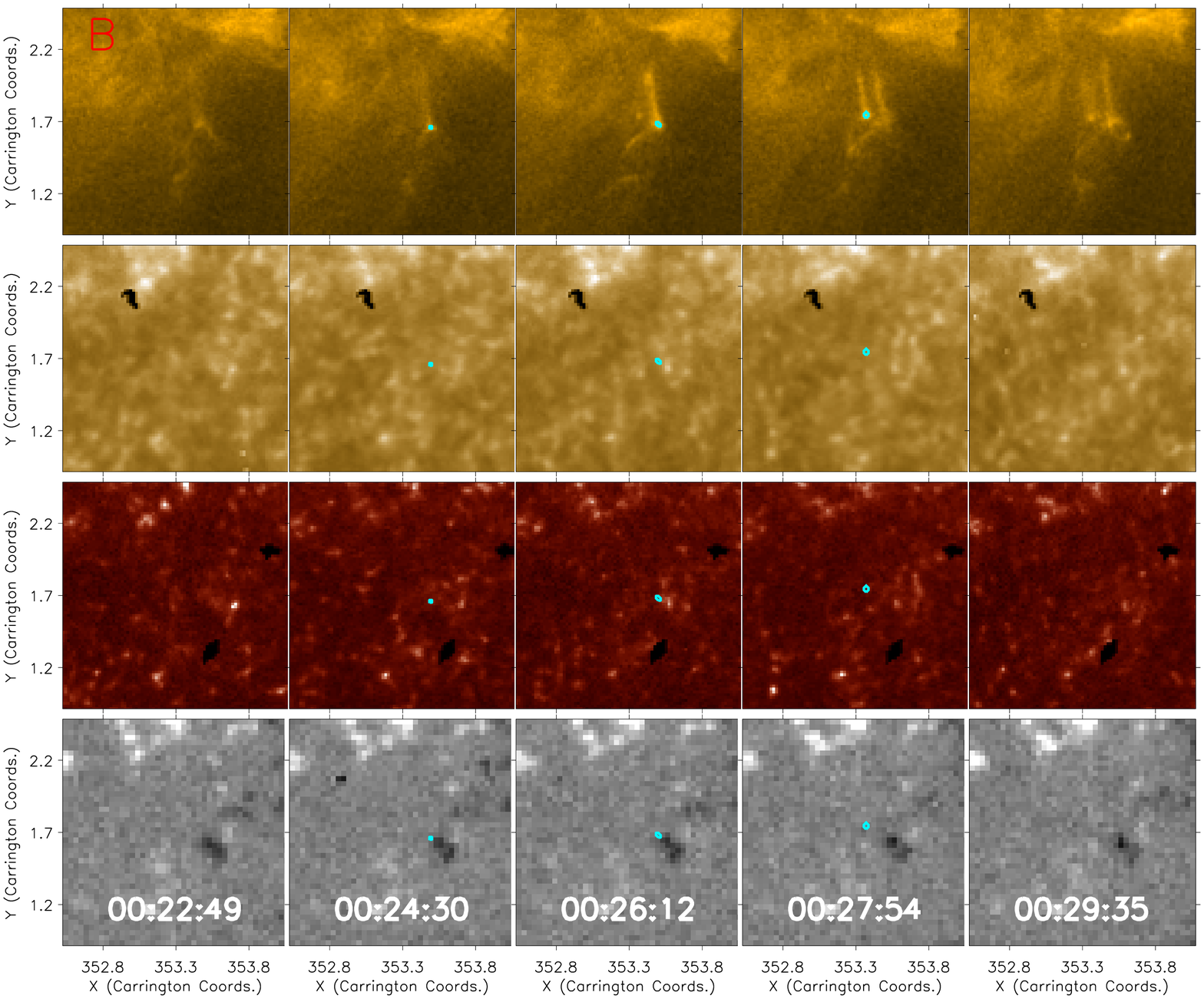}
\caption{Same as for Fig.~\ref{CFs_IRIS_F4} but for an EUV brightening which has no signature in either IRIS SJI filter. A clear structure is evident in the \HRI\ images, with the automatically detected EUV brightening (outlined by the cyan contours) being detected where two almost perpendicular arms appear to meet. Neither the arms nor a transient point-like brightening at the location of the EUV brightening are apparent in the IRIS SJI filter data. An animation of this region is available with the online version of this article.}
\label{CFs_IRIS_B2}
\end{figure*}

In the previous subsections, we analysed four regions (`D'-`G') of EUV brightening activity where clear burst-like signatures were also present in IRIS imaging and, for one example, spectral data. We note, however, that this is not a one-to-one relationship and that a number of EUV brightenings do not appear to display any transition region response. Specifically within this dataset, we identified four periods of EUV brightening activity within regions `A'-`C' which displayed no discernible response in the IRIS SJI data. In Fig.~\ref{CFs_IRIS_B2}, we plot a clear example of an automatically detected EUV brightening, contained within a larger structure, for which no IRIS response was detected throughout its lifetime. The top row plots \HRI\ images around the event with a cadence of approximately $101$ s. Over these seven minutes, a long and thin structure with two perpendicular arms appears in the \HRI\ images (most clearly seen in the third column), with an EUV brightening being detected where the arms appear to meet (cyan contour). By the fourth column, a second vertically orientated arm had appeared with a second EUV brightening being detected at the base of this. Neither the distinct arms of the larger structure nor the EUV brightening activity detected in the \HRI\ data were discernible in the IRIS \ion{Mg}{II} $279.6$ nm or \ion{Si}{IV} $140$ nm filters (second and third rows, respectively). Additionally, the line-of-sight magnetic field did not display any clear evidence of a bi-pole nor cancellation at the location of this event (bottom row).

The other three example instances of EUV brightening activity within regions `A'-`C' which displayed no clear response in co-spatial and co-temporal IRIS imaging data were similar to the example plotted in Fig.~\ref{CFs_IRIS_B2}, in that they were point-like regions (diameters of several pixels) with lifetimes of between one and four minutes. It is possible that such events correspond to energy release in the upper solar atmosphere, where the plasma is already at temperatures above those sampled by the IRIS filters prior to the additional input of energy. However, given that these IRIS SJI data are summed spatially (i.e. the spatial sampling is double the highest possible resolution), it is also possible that an IRIS response would have been present in higher resolution data, but that it is unresolved here. This would need to be tested through examination of unsummed IRIS SJI data collected during subsequent coordinated observations. Overall, we are not in a position to estimate the percentage of EUV brightenings which display an IRIS response due to our small sample size, but we are able to confirm that this value is non-zero. Movies depicting the EUV brightening activity in regions `A'-`C', as well as the IRIS SJI \ion{Mg}{II} $279.6$ nm and \ion{Si}{IV} $140$ nm filter intensities, are included with the online version of this article.

\section{Discussion}
\label{Discussion}

The discovery of EUV brightenings on extremely small-scales (\citealt{Berghmans21}), and their apparent links to magnetic reconnection (e.g. \citealt{Chen21,Kahil22}), has been one of the key early results of Solar Orbiter. In this article, we investigated whether signatures of EUV brightenings were evident in transition region imaging and spectral data sampled by the IRIS satellite. Through our analysis of seven distinct spatial regions (labelled `A'-`G'), we found that some but not all EUV brightenings display co-spatial and co-temporal responses in IRIS imaging data. Due to the varied and complex nature of the results obtained through our analysis, we provide a brief overview here and a summary in Table~\ref{Tab_overview}.

In regions `D'-`G', we found seven instances of extended (long and thin structures; see Figs.~\ref{CFs_IRIS_E1}-\ref{CFs_IRIS_F4}) EUV brightening activity and four instances of more complex (rapidly evolving structures; see Fig.~\ref{CFs_IRIS_F3}) EUV brightening activity co-spatial to brightenings in data sampled by the \ion{Si}{IV} $140$ nm SJI filter. Eight of these events also displayed signatures in the \ion{Mg}{II} $279.6$ nm filter suggesting that EUV brightenings can occur across a range of temperature windows. In regions `A'-`C', however, we found four instances of point-like EUV brightenings which displayed no co-spatial and co-temporal intensity increases in IRIS SJI data (see Fig.~\ref{CFs_IRIS_B2}). Whether unsummed (in the spatial domain) IRIS SJI data would have returned a signature needs to be tested through analysis of data sampled during other coordinated observations. Our small sample size does not allow us to infer any estimate for the overall number of EUV brightenings which display signatures in transition region imaging, but we are able to conclude that this number is non-zero.

Analysis of lightcurves, constructed from both \HRI\ and IRIS SJI data, sampled co-spatial to EUV brightenings which displayed intensity increases in both, returned contrasting behaviour. For three events (both extended and complex), intensity increases were apparent in the IRIS imaging data between $100$-$150$ s prior to the EUV brightening related intensity increase in the \HRI\ data, suggesting heating of the plasma from temperatures below $8\times10^{4}$ K (IRIS SJI \ion{Si}{IV} $140$ nm filter temperature response) to above $8\times10^{5}$ K (\HRI\ temperature response). For two events, however, the intensity increases occurred co-temporally, potentially implying a single instance of energy release across multiple temperature ranges. Finally, for one example (plotted in Fig.~\ref{CFs_IRIS_E1}) the \HRI\ intensity increase occurred prior to the co-spatial brightening in the \ion{Si}{IV} imaging data, potentially indicating that plasma cooling from temperatures above $8\times10^{5}$ K to temperatures below $8\times10^{4}$ K could be causing the EUV brightening signature. Lightcurves constructed from the other EUV brightenings were inconclusive. These results imply that EUV brightenings (as returned by the algorithm) are not always evidence of current heating events, with some potentially pin-pointing locations of plasma cooling through the \HRI\ temperature passband. This could occur for several reasons including, but not limited to, previous heating at the same location which may have been too impulsive to be detected by instruments (\citealt{Cargill94, Cargill15, Parenti19}) or the occurrence of the thermal instability in a coronal loop leading to the detection of rapidly cooling plasma at the loop foot-points (\citealt{Nelson20, Nelson20b}). We favour this second interpretation for the event reported in this article. The lightcurves discussed here can be found in Fig.~\ref{CFs_IRIS_lightcurves} and Fig.~\ref{CFs_IRIS_lightcurves_exp}.

Finally, we also analysed the spectral response of several lines sampled by IRIS co-spatial to an EUV brightening which occurred along the slit (see Figs.~\ref{CFs_IRIS_F4}, \ref{CFs_IRIS_Spec}-\ref{CFs_IRIS_SpT}). Both intensity increases and spectral broadening were evident in all of the \ion{Mg}{II} k $279.6$ nm, \ion{C}{II} $133.7$ nm, \ion{Si}{IV} $139.4$ nm, and \ion{Si}{IV} $140.3$ nm lines co-spatial to this event. Additionally, the ratio between the \ion{Si}{IV} $139.4$ nm and $140.3$ nm lines was around $1.7$ indicating this event may have been formed under optically thick conditions, whilst oppositely directed Doppler shifts, with speeds of around $22$ km s$^{-1}$ were identified at either end of the EUV brightening. These combined results indicate that the transition region brightening co-spatial to the EUV brightening may be an EE (as discussed by for example \citealt{Teriaca04, Huang14,Huang19}). Unfortunately, no density estimates were possible due to the short exposure times returning low \ion{O}{IV} intensities.

\section{Conclusions}
\label{Conclusions}

The situation we find here in terms of the thermal connectivity is somewhat reminiscent of that found previously for other similar phenomena. For example, spectroscopic time series observations in the temperature minimum, chromosphere and transition region show a comparably complex picture in that sometimes brightenings can be seen in purely one temperature domain, or simultaneously across two or three of these distinct temperature regions (\citealt{Brkovic03}). A clear example of this is how UV bursts and EBs can both occur individually or together (e.g. \citealt{Vissers15,Chen19}). In EBs strong horizontal flows push together vertical magnetic field of opposite direction (in regions of high plasma-beta) leading to very efficient reconnection low down in the solar atmosphere (\citealt{Hansteen17,Danilovic17}). The co-formation of EBs and UV bursts can be explained by reconnection occurring at opposite ends of a long current sheet in response to emerging magnetic flux (\citealt{Hansteen19}). For the events observed here (which appear to occur higher in the atmosphere), however, the question arises as to what might determine the different response of the plasma, in terms of transition region or coronal emission, to magnetic heating. EEs are typically modelled through reconnection of oppositely directed magnetic field in a low-beta plasma (see, for example, \citealt{Innes99}) which could hold for the EUV brightenings observed here. This raises the question of the role of plasma-beta for these reconnection events (as discussed by \citealt{Peter19}), and if one might be able to find a unified reconnection picture for all these events and their thermal response at different temperatures.

Overall, our results indicate that there is no `typical' response to EUV brightenings in the transition region, at least as observed by the IRIS imagers. Some events display clear co-spatial and co-temporal signatures, often with similar morphological properties, in \ion{Mg}{II} and \ion{Si}{IV} imaging data whereas others display no detectable response at all. The single EUV brightening which could be confidently identified along the IRIS slit had spectral signatures comparable to EEs in the transition region (see, for example, \citealt{Teriaca04}). Work analysing additional co-observations with IRIS sampled during future Solar Orbiter Remote Sensing Windows will allow us to make further inferences about the potential links between EUV brightenings and transition region events in the future.

\begin{table*}[!hbt]
\centering
\caption{Summary of the $15$ EUV brightenings studied in this article.}
\begin{tabular}{| c | c | c | c | c | c | c | c | c | c | c |}
\hline
\bf{Region} & \bf{Event} & \bf{Type} & \bf{\ion{Mg}{II}} & \bf{\ion{Si}{IV}} & \bf{Co-spatial} & \bf{Spectra} & \bf{Feature} & \bf{T$_\mathrm{START}$ (UT)} & \bf{T$_\mathrm{END}$ (UT)} & \bf{Bi-pole} \\ \hline
A & 1 & Point & No & No & - & - & - & $00$:$22$:$28$ & $00$:$26$:$32$ & No \\ \hline
B & 1 & Point & No & No & - & - & - & $00$:$08$:$14$ & $00$:$12$:$18$ & Yes \\ \hline
B & 2 & Point & No & No & - & - & - & $00$:$22$:$49$ & $00$:$29$:$35$ & No \\ \hline
C & 1 & Point & No & No & - & - & - & $00$:$27$:$03$ & $00$:$29$:$45$ & No \\ \hline
D & 1 & Extended & No & Yes & Yes & - & - & $00$:$12$:$08$ & $00$:$15$:$31$ & No \\ \hline
D & 2 & Extended & No & Yes & Yes & - & - & $00$:$20$:$16$ & $00$:$23$:$39$ & No \\ \hline
D & 3 & Extended & No & Yes & Yes & - & - & $00$:$25$:$31$ & $00$:$27$:$33$ & No \\ \hline
E & 1 & Extended & No & Yes & No & - & - & $00$:$09$:$05$ & $00$:$12$:$29$ & No \\ \hline
E & 2 & Extended & Yes & Yes & Yes & - & - & $00$:$15$:$21$ & $00$:$18$:$45$ & No \\ \hline
F & 1 & Complex & Yes & Yes & - & - & - & $00$:$04$:$20$ & $00$:$08$:$24$ & Yes \\ \hline
F & 2 & Extended & Yes & Yes & Yes & - & - & $00$:$14$:$20$ & $00$:$16$:$22$ & Yes \\ \hline
F & 3 & Complex & Yes & Yes & - & - & - & $00$:$16$:$22$ & $00$:$22$:$28$ & Yes \\ \hline
F & 4 & Extended & Yes & Yes & Yes & Yes & EE & $00$:$23$:$39$ & $00$:$29$:$45$ & No \\ \hline
G & 1 & Complex & Yes & Yes & - & - & - & $00$:$04$:$20$ & $00$:$07$:$44$ & Yes \\ \hline
G & 2 & Complex & Yes & Yes & - & - & - & $00$:$12$:$49$ & $00$:$18$:$55$ & Yes \\ \hline
\end{tabular}

\tablefoot{General information about the $15$ EUV brightenings studied here, including: The region in which they were identified; their label within that region; the type of EUV brightening; whether there was a response in the IRIS \ion{Mg}{II} SJI data; whether there was a response in the IRIS \ion{Si}{IV} SJI data; whether this response was co-spatial to the EUV brightening; whether IRIS spectra were available for the event; the type of spectra returned; the start of the temporal region of interest around the event; the end of the temporal region of interest around the event; and whether a (tentative) bi-pole was apparent in SDO/HMI data.}
\label{Tab_overview}
\end{table*}

\begin{acknowledgements}
CJN is thankful to ESA for support as an ESA Research Fellow. Solar Orbiter is a mission of international cooperation between ESA and NASA, operated by ESA. DML is grateful to the Science Technology and Facilities Council for the award of an Ernest Rutherford Fellowship (ST/R003246/1). SP acknowledges the funding by CNES through the MEDOC data and operations center. The EUI instrument was built by CSL, IAS, MPS, MSSL/UCL, PMOD/WRC, ROB, LCF/IO with funding from the Belgian Federal Science Policy Office (BELSPO/PRODEX PEA 4000134088, 4000112292, 4000136424, and 4000134474); the Centre National d’Etudes Spatiales (CNES); the UK Space Agency (UKSA); the Bundesministerium f\"ur Wirtschaft und Energie (BMWi) through the Deutsches Zentrum f\"ur Luft und Raumfahrt (DLR); and the Swiss Space Office (SSO). IRIS is a NASA small explorer mission developed and operated by LMSAL with mission operations executed at NASA Ames Research Center and major contributions to downlink communications funded by ESA and the Norwegian Space Centre. SDO/AIA and SDO/HMI data provided courtesy of NASA/SDO and the AIA and HMI science teams. This research has made use of NASA’s Astrophysics Data System Bibliographic Services.
\end{acknowledgements}

\bibliographystyle{aa}
\bibliography{SOLO_IRIS}

\end{document}